\documentstyle[graphicx,12pt]{article}
\begin{document}

\small
\hoffset=-1truecm
\voffset=-2truecm
\title{\bf The virial relation for compact Q-balls in the complex signum-Gordon model}
\author{Huawen Wang \hspace {1cm} Hongbo Cheng\footnote {E-mail address:
hbcheng@sh163.net}\\
\footnote{021-64703389(H);13661864435(M)}
Department of Physics,
East China University of Science and
Technology,\\ Shanghai 200237, China\\
The Shanghai Key Laboratory of Astrophysics,\\ Shanghai 200234,
China}

\date{}
\maketitle

\begin{abstract}
In this work the properties of Q-balls in the complex
signum-Gordon model in $d$ spatial dimensions is studied. We
obtain a general virial relation for this kind of Q-ball in the
higher-dimensional spacetime. We compute the energy and radii of
Q-ball with V-shaped field potential as a function of spatial
dimensionality and a parameter defining the model potential energy
density to show that this kind of Q-balls can also be survive
stably in the high-dimensional spacetime.
\end{abstract}
\vspace{8cm} \hspace{1cm} PACS number(s): 11.27.+d, 11.10.Lm,
98.80.Cq

\newpage

The nontopological solitons possess a conserved Noether charge
because of a symmetry of the Lagrangian of system in contrast to
the topological charge resulting from the spontaneous symmetry
breaking in the case of topological defects [1, 2]. As
nontopological solitons Q-balls appear in extended localized
solutions of models with certain self-interacting complex scalar
field [3]. Their stability is associated with their charge $Q$ and
that their mass is smaller than the mass of a collection of scalar
fields. The unbroken continuous globally symmetry like $U(1)$ is
also possessed. The standard models have the smooth potentials
near their absolute minima, so they hold exponential tails and
interact with each other. Such solitons are physically universal
and have been studied in several subjects like dark matter [4-6],
condensed matter physics [7] and so on. The Q-balls have also
attracted much attentions in cosmology [8-11]. The Boson stars can
be considered as solutions of complex scalar field models couple
to gravity [12-19]. The general nontopological solitons have also
been explored in de Sitter and anti de Sitter spacetimes
respectively [20, 21].

There is plenty of motivation to research on the nontopological
soliton models in the higher-dimensional spacetimes. The issue of
higher-dimensional spacetime can help us to unify the interactions
in nature with extra compactified dimensions and to solve the
hierarchy problem with an additional warped dimension [22-25]. The
signatures of these extra dimensions may be explored in various
directions. The properties of high-dimensional world also dominate
the structure and the stability of various kinds of nontopological
soliton models including Q-balls. We can search for the dependence
of Q-balls configuration and energy subject to a finite charge on
the spatial dimensions. Q-balls may survive well in a
precompactification world while modify the structure of the
vacuum. Studying the Q-balls in the Universe with more than four
dimensions certainly help us to probe the background with
additional spatial dimensions.

A new kind of interesting nontopological solitons was put forward
[26] and some important efforts have been contributed [27]. The
so-called complex signum-Gordon model has the field potential with
sharp bottom rather than wholely smooth one. Further the V-sharped
field potential can be denoted as $U(\phi)=\lambda|\phi|$, where
$\lambda$ is positive constant and $\phi$ is a complex scalar
field. The inverted-conical-shaped potentials are certainly
different from the case of other nontopological solitons with
potential involving higher order of fields. There exist the
Q-balls in the complex signum-Gordon model, which are stationary
and whose energy is finite while their scalar field, energy
density and charge density approach to the zero outside the ball
region. It should be pointed out that these Q-ball solutions are
simpler than the others. The boson stars and black holes in scalar
electrodynamics with a V-shaped scalar potential are considered
[28]. The spacetime around this kind of boson stars consists of a
Schwarzschild-type black hole in the interior and a
Reissner-Nordstrom-type spacetime in the exterior appears.

It is significant to study the compact Q-balls in the complex
signum-Gordon model in higher-dimensional spacetimes by means of
virial theorem. The task for exploring the Q-balls with V-shaped
field potential in the four-dimensional background [26-28], but
the properties of this kind of Q-balls in the higher-dimensional
spacetimes have not studied in detail. A generalized virial
relation for Q-balls involving general smooth potentials like
$V(\phi\phi^{+})=\sum_{n=1}^{\infty}a_{n}(\phi\phi^{+})^{n}$ with
constants $a_{n}$ and integer $n$ in $d$ spatial dimensions was
obtained and how spatial dimensionality affects some of the key
properties of Q-balls such as their energy, minimal charge and
size was also declared [29]. We will study the same kind of
Q-balls in the high-dimensional spaacetime by means of the method
from Ref. [29]. We should utilize the virial theorem to bring
about the necessary conditions on the complex signum-Gordon model
while find the model's conserved charge $Q$ and energy associated
with the dimensionality and the parameters in the potential.

In this paper we discuss the Q-balls in the complex signum-Gordon
model in the world with arbitrary dimensions carefully. We present
a d-dimensional virial relation for this kind of Q-balls and the
relation will recover to be the Derrick's theorem when $Q=0$. We
study these Q-balls to estimate their properties in the case of
large charge and radius and small ones respectively. Our results
are listed in the end.

Here we consider the complex signum-Gordon model with Lagrangian
density as follow,

\begin{equation}
{\mathcal{L}}=\partial_{\mu}\phi^{+}\partial^{\mu}\phi-V(\phi\phi^{+})
\end{equation}

\noindent where $\phi$ is a complex scalar field in
($d+1$)-dimensional Minkowski spacetime. The index
$\mu=0,1,2,\cdot\cdot\cdot,d$ and the signature is $(+, -, -,
\cdot\cdot\cdot)$. The potential is assumed to be
$V(\phi\phi^{+})=\lambda|\phi|$ with a global minimum at $\phi=0$
and the coupling constant $\lambda>0$. As Q-balls this model is
nonperturbative excitation about this global vacuum state carring
a net particle number named charge $Q$ which is conserved. The
condition that the energy of the Q-ball $E_{Q}$ is smaller than
$Qm_{\phi}$ which is energetically preferred to keep its own
stability. The Lagrangian of the complex signum-Gordon model has a
conserved $U(1)$ symmetry under the global transformation
$\phi(x)\longrightarrow e^{i\alpha}\phi(x)$. The associated
conserved current should be
$j^{\mu}=-i(\phi^{+}\partial^{\mu}\phi-\phi\partial^{\mu}\phi^{+})$
and the corresponding conserved charge is given by $Q=\int
d^{d}xj^{0}$. We introduce the ansatz for configurations with
lowest energy,

\begin{equation}
\phi(x)=\frac{1}{\sqrt{2}}\Phi{(\mathbf{x})}e^{i\omega t}
\end{equation}

\noindent Here the field $\Phi(\mathbf{x})$ can be taken to be
spherically symmetry and $\{\mathbf{x}\}$ represent the spatial
components of coordinates. The field equation can be obtained,

\begin{equation}
(\nabla_{d}^{2}+\omega^{2})\Phi-\frac{\lambda}{2}\frac{\Phi}{|\Phi|}=0
\end{equation}

\noindent According to Ref. [29] and from Lagrangian density (1)
and ansatz (2), the energy for a field configure in a
($d+1$)-dimensional spacetime can be obtained,

\begin{equation}
E[\Phi]=\int d^{d}x[\frac{1}{2}(\nabla_{d}\Phi)^{2}
+\frac{1}{2}\omega^{2}\Phi^{2}+V(\Phi^{2})]
\end{equation}

\noindent We can also denote the energy in terms of the conserved
charge $Q$ as follow,

\begin{equation}
E_{Q}[\Phi]=\frac{1}{2}\frac{Q^{2}}{\langle\Phi^{2}\rangle}
+\langle\frac{1}{2}(\nabla_{d}\Phi)^{2}\rangle+\langle V\rangle
\end{equation}

\noindent where $\langle\cdot\cdot\cdot\rangle=\int\cdot\cdot\cdot
d^{d}x$. In order to obtain the virial relation we scale the
spatial variable $x\longrightarrow\alpha x$ and impose the
invariance of energy under scale transformation $\frac{\partial
E}{\partial\alpha}|_{\alpha=1}=0$. As a generalization of
Derrick's theorem the virial relation for Q-balls in a spacetime
with arbitrary dimensionality and written as,

\begin{equation}
d\langle
V\rangle=(2-d)\langle\frac{1}{2}(\nabla_{d}\Phi)^{2}\rangle+\frac{d}{2}
\frac{Q^{2}}{\langle\Phi^{2}\rangle}
\end{equation}

\noindent It should be pointed that the Derrick's theorem can be
recovered for $Q=0$. Further the absolute lower bound for Q-balls
to be a preferred energy state is shown as
$Q^{2}\geq\frac{2(d-2)}{d}\langle\Phi^{2}\rangle
\langle\frac{1}{2}(\nabla_{d}\Phi)^{2}\rangle$ because of $\langle
V\rangle\geq 0$. Combining Eq. (4)-(6), we express the energy
density as,

\begin{equation}
\frac{E}{Q}=\omega(1+\frac{1}{d-2+d\frac{\langle V\rangle}
{\langle\frac{1}{2}(\nabla_{d}\Phi)^{2}\rangle}})\leq m_{\phi}
\end{equation}

\noindent Only the Q-balls satisfying the condition (7) are stable
instead of dispersing. Here we choose,

\begin{equation}
\langle V\rangle=\frac{\lambda}{\sqrt{2}}\int d^{d}x|\Phi|
\end{equation}

\noindent then the formulae mentioned above can be utilized to
explore the compact Q-balls of complex signum-Gordon model in
higher-dimensional world analytically and the nonlinear field
equation needs not be solved.

Now in $(d+1)$-dimensional spacetimes we work on the
complex-signum-Gordon-type Q-balls which are characterized by
large charge and radius. At first we utilize the Coleman issue [3]
to explore the large Q-balls. We choose the field to be a step
function which is equal to be a constant denoted as $\Phi_{c}$
within the ball and vanishes out side the ball's volume $v$. The
step-function type field leads
$\langle\Phi^{2}\rangle=\Phi_{c}^{2}v$, $\nabla_{d}\Phi=0$ and
$\langle V\rangle=\frac{\lambda}{2}|\Phi_{c}|v$. From Eq.(5) the
model energy is

\begin{equation}
E=\frac{1}{2}\frac{Q^{2}}{\Phi_{c}^{2}v}+\frac{\lambda}{\sqrt{2}}|\Phi_{c}|v
\end{equation}

\noindent Having extremized the expression (9) with respect to the
volume $v$, we obtain the minimum energy and the condition for
Q-ball's stability as follow,

\begin{equation}
\frac{E_{\min}}{Q}=2^{\frac{1}{4}}(\frac{\lambda}{\Phi_{c}})^{\frac{1}{2}}
<m_{\phi}
\end{equation}

\noindent The complex signum-Gordon Q-ball can exist if the model
parameters are adjusted reasonably.

In order to describe the large Q-balls in the complex
signum-Gordon model, we introduce the field profile,

\begin{eqnarray}
\Phi(r)=\{\begin{array}{cc}
  \Phi_{c} & r<R \\
  \Phi_{c}e^{-\alpha(r-R)} & r\geq R \\
\end{array}
\end{eqnarray}

\noindent where $\alpha$ is a variational parameter. This kind of
function keeps the scalar field to be a constant like $\Phi_{c}$
within an area and not to drop to the zero in the larger region
whose size is larger than $R$. According to large-Q-ball ansatz
(11), the energy of model reads,

\begin{eqnarray}
E=\frac{1}{2}\frac{Q^{2}}{\langle\Phi^{2}\rangle}
+\frac{1}{2}\langle(\nabla_{d}\Phi)^{2}\rangle
+\frac{\lambda}{\sqrt{2}}\langle|\Phi|\rangle\hspace{2.5cm}\nonumber\\
=\frac{1}{2}\omega^{2}\frac{c_{d}}{d}\Phi_{c}^{2}R^{d}
+\frac{1}{2}\omega^{2}c_{d}\Phi_{c}^{2}(d-1)!
\sum_{k=0}^{d-1}\frac{1}{k!}\frac{1}{(2\alpha)^{d-k}}R^{k}\nonumber\\
+\frac{1}{2}c_{d}\Phi_{c}^{2}\alpha^{2}(d-1)!
\sum_{k=0}^{d-1}\frac{1}{k!}\frac{1}{(2\alpha)^{d-k}}R^{k}\hspace{2cm}\nonumber\\
+\frac{\lambda}{\sqrt{2}}\frac{c_{d}}{d}\Phi_{c}R^{d}
+\frac{\lambda}{\sqrt{2}}c_{d}\Phi_{c}(d-1)!
\sum_{k=0}^{d-1}\frac{1}{k!}\frac{1}{\alpha^{d-k}}R^{k}
\end{eqnarray}

\noindent while the conserved charge is obtained,

\begin{eqnarray}
Q=\omega\int d^{d}x\Phi^{2}\hspace{5cm}\nonumber\\
=\omega\frac{c_{d}}{d}\Phi_{c}^{2}R^{d}+\omega
c_{d}\Phi_{c}^{2}(d-1)!\sum_{k=0}^{d-1}\frac{1}{k!}\frac{1}{(2\alpha)^{d-k}}R^{k}
\end{eqnarray}

\noindent which is the same as that of Ref. [29], here

\begin{equation}
c_{d}=\frac{2\pi^{\frac{d}{2}}}{\Gamma(\frac{d}{2})}
\end{equation}

\noindent The conserved charge is able to replace the frequency
$\omega$. We just keep the dominant terms in the expressions for
the energy and this approximation is accurate enough for large
Q-balls. Combining the expression for $Q$ in Eq. (13), we find

\begin{equation}
E_{Q}\leq\frac{Q^{2}d}{2c_{d}\Phi_{c}^{2}}R^{-d}
+\frac{c_{d}b}{\sqrt{2}d}R^{d}+(\frac{c_{d}\alpha\Phi_{c}^{2}}{4}
+\frac{c_{d}b}{\sqrt{2}\alpha})R^{d-1}
\end{equation}

\noindent where

\begin{equation}
b=\lambda\Phi_{c}
\end{equation}

It is clear that the model energy depends on the variables $R$ and
$\alpha$ not belonging to the model parameters. We extremise the
energy expression (15) with respect to $R$ and $\alpha$
respectively. By means of performance $\frac{\partial E}{\partial
R}|_{R=R_{c}}=0$ we find the equation that the critical radius
$R_{c}$ satisfies,

\begin{equation}
-\frac{Q^{2}d^{2}}{2c_{d}\Phi_{c}^{2}}+\frac{c_{d}b}{\sqrt{2}}R^{2d}
+(d-1)(\frac{c_{d}\Phi_{c}^{2}\alpha}{4}+\frac{c_{d}b}{\sqrt{2}\alpha})
R^{2d-1}=0
\end{equation}

\noindent The approximate solution to Eq. (17) is,

\begin{equation}
R_{c}\approx2^{-\frac{1}{4d}}(\lambda\Phi_{c})^{-\frac{1}{2d}}
(\frac{Qd}{c_{d}\Phi_{c}})^{\frac{1}{d}}
-\frac{d-1}{\sqrt{2}d}(\frac{\alpha\Phi_{c}}{4\lambda}+\frac{1}{\sqrt{2}\alpha})
\end{equation}

\noindent and the solution is valid for large Q-balls. We can also
impose the condition $\frac{\partial
E_{Q}}{\partial\alpha}|_{\alpha=\alpha_{c}}=0$ into Eq. (15) to
find,

\begin{equation}
\alpha_{c}=2^{\frac{3}{4}}(\frac{\lambda}{\Phi_{c}})^{\frac{1}{2}}
\end{equation}

\noindent Substituting the results (18) and (19) into Eq. (15), we
obtain the extremised expression for the energy of large Q-ball,

\begin{equation}
\frac{E_{Q}[\Phi_{c}]|_{R_{c},\alpha_{c}}}{Q}
\approx2^{\frac{1}{4}}(\frac{\lambda}{\Phi_{c}})^{\frac{1}{2}}
(1+\xi_{c}Q^{-\frac{1}{d}})
\end{equation}

\noindent where

\begin{equation}
\xi_{c}=2^{\frac{1}{4}(\frac{1}{d}-3)}c_{d}^{\frac{1}{d}}
(\frac{d}{\sqrt{\lambda}})^{1-\frac{1}{d}}\Phi_{c}^{\frac{1}{2}(1+\frac{3}{d})}
\end{equation}

\noindent Having compared the results in Eq. (20) with those in
Eq. (10), we also declare that the lower bound on the energy per
one particle in large Q-balls is larger than that of Coleman
approach. It should be pointed out that
$\frac{E_{Q}[\Phi_{c}]|_{R_{c},\alpha_{c}}}{Q}$ is finite within
the context of large radius and conserved charge. We plot the
dependence of the minimum energy per unit charge of
complex-signum-Gordon-type Q-ball on the charge with $\lambda=1$
and $\Phi_{c}=1$ for different spatial dimensions in Fig. 1 and
show that the minimum energy per unit charge is a decreasing
function of the charge. As the charge $Q$ becomes extremely large,
$\frac{E_{Q}[\Phi_{c}]|_{R_{c},\alpha_{c}}}{Q}$ will approach to
the value which is just associated with the model parameters no
matter how high the dimensionality is. The Q-ball will possess the
larger $\frac{E_{Q}[\Phi_{c}]|_{R_{c},\alpha_{c}}}{Q}$ in the
higher-dimensional world.

We start to discuss the small Q-balls in the signum-Gordon model
in the $(d+1)$-dimensional spacetimes. The small Q-balls with
radii $R\geq m_{\phi}^{-1}$ can not be described well by means of
thin-wall approximation. We introduce a Gaussian ansatz in order
to consider the small Q-balls in the complex signum-Gordon model,

\begin{equation}
\Phi({\mathbf{x}})=\Phi_{c}e^{-\frac{r^{2}}{R^{2}}}
\end{equation}

\noindent Substituting the ansatz (22) into the expression (5), we
obtain the small Q-ball energy as follow,

\begin{eqnarray}
E=\int
d^{d}x[\frac{1}{2}\omega^{2}\Phi^{2}+\frac{1}{2}(\nabla_{d}\Phi)^{2}
+V(\Phi^{2})]\hspace{1.5cm}\nonumber\\
=(\frac{\pi}{2})^{\frac{d}{2}}[\frac{1}{2}(\frac{2}{\pi})^{d}
\frac{Q^{2}}{\Phi_{c}^{2}}R^{-d}+\frac{d}{2}\Phi_{c}^{2}R^{d-2}
+2^{\frac{d-2}{2}}\lambda|\Phi_{c}|R^{d}]
\end{eqnarray}

\noindent The energy for small Q-ball in the signum-Gordon model
just contains the dominant terms. By extremizing the expression
for the energy with respect to $R$ like $\frac{\partial
E_{Q}}{\partial R}|_{R=R_{c}}=0$ we obtain the equation for the
critical radius $R_{c}$ as,

\begin{equation}
R_{c}^{2d}+\frac{d-2}{d}\frac{d}{2^{\frac{d+1}{2}}}\frac{\Phi_{c}}{\lambda}
R_{c}^{2d-2}-\frac{2^{\frac{d-1}{2}}}{\pi^{d}}\frac{Q^{2}}{\lambda}
\frac{1}{|\Phi_{c}|^{3}}=0
\end{equation}

\noindent Similarly the approximate solution reads,

\begin{equation}
R_{c}\approx R_{0}(1-\frac{d-2}{d}\frac{1}{2^{\frac{d+3}{2}}}
\frac{\Phi_{c}}{\lambda}\frac{1}{R_{0}^{2}})
\end{equation}

\noindent where

\begin{equation}
R_{0}=(\frac{2^{\frac{d-1}{2}}}{\pi^{d}}\frac{Q^{2}}{\lambda|\Phi_{c}|^{3}})^{\frac{1}{2d}}
\end{equation}

\noindent According to the critical radius the energy can be
expressed as

\begin{equation}
\frac{E[\Phi_{c}]|_{R=R_{c}}}{Q}=2^{\frac{d+1}{4}
}(\frac{\lambda}{\Phi_{c}})^{\frac{1}{2}}
[1+\frac{\eta_{c}Q^{-\frac{2}{d}}}{2}-\frac{(d-2)^{2}}{4d^{2}}
\eta_{c}^{2}Q^{-\frac{4}{d}}]
\end{equation}

\noindent where

\begin{equation}
\eta_{c}=\frac{d}{2^{\frac{d}{2}-\frac{1}{2d}+1}}\pi
\lambda^{\frac{1-d}{d}}\Phi_{c}^{\frac{3}{d}+1}
\end{equation}

\noindent The stability of small Q-balls can be held according to
Eq. (27). The energy over charge is finite while the model
parameters can be adjusted. The powers of charge $Q$ in the terms
are negative, so the model energy per unit charge will not be
divergent as the charge becomes larger. In Fig. 2, for simplicity
we also choose $\lambda=1$ and $\Phi_{c}=1$ without losing
generality and we show that the minimum energy per unit charge
$\frac{E[\Phi_{c}]|_{R=R_{c}}}{Q}$ also decreases as the charge
increases in the case of small charge and radius in the background
with different spatial dimensions. The higher dimensionality will
lead the larger $\frac{E[\Phi_{c}]|_{R=R_{c}}}{Q}$. Certainly here
the energy over charge is smaller than that in the case of large
ones.

In this work we investigate the compact Q-balls in the complex
signum-Gordon model in the spacetime with arbitrary
dimensionality. The Q-ball properties associated with the
background structure may help us to further investigate our world.
This kind of nontopological model is simple but is enough to
contain the properties of Q-balls. We search for the approximate
and acceptable analytical description for characteristics of the
compact Q-balls in the complex signum-Gordon model instead of
solving the nonlinear field equations numerically. We find the
virial relation for these kinds of Q-balls. We study the model
with V-shaped field potential in the case of large Q-balls and
small ones. We obtain the approximate analytical expressions to
show the Q-balls' energy and radii depending on the spatial
dimensionality and the parameters belonging to the complex
signum-Gordon model. The minimum energy per unit charge of the
system decreases to a quantity depending on the system parameters
and dimensionality. It should be pointed out that the accuracy of
the approximate analytical expressions can be acceptable.

\vspace{3cm}

\noindent\textbf{Acknowledgement}

This work is supported by NSFC No. 10875043 and is partly
supported by the Shanghai Research Foundation No. 07dz22020.

\newpage
\begin{figure}
\setlength{\belowcaptionskip}{10pt} \centering
  \includegraphics[width=15cm]{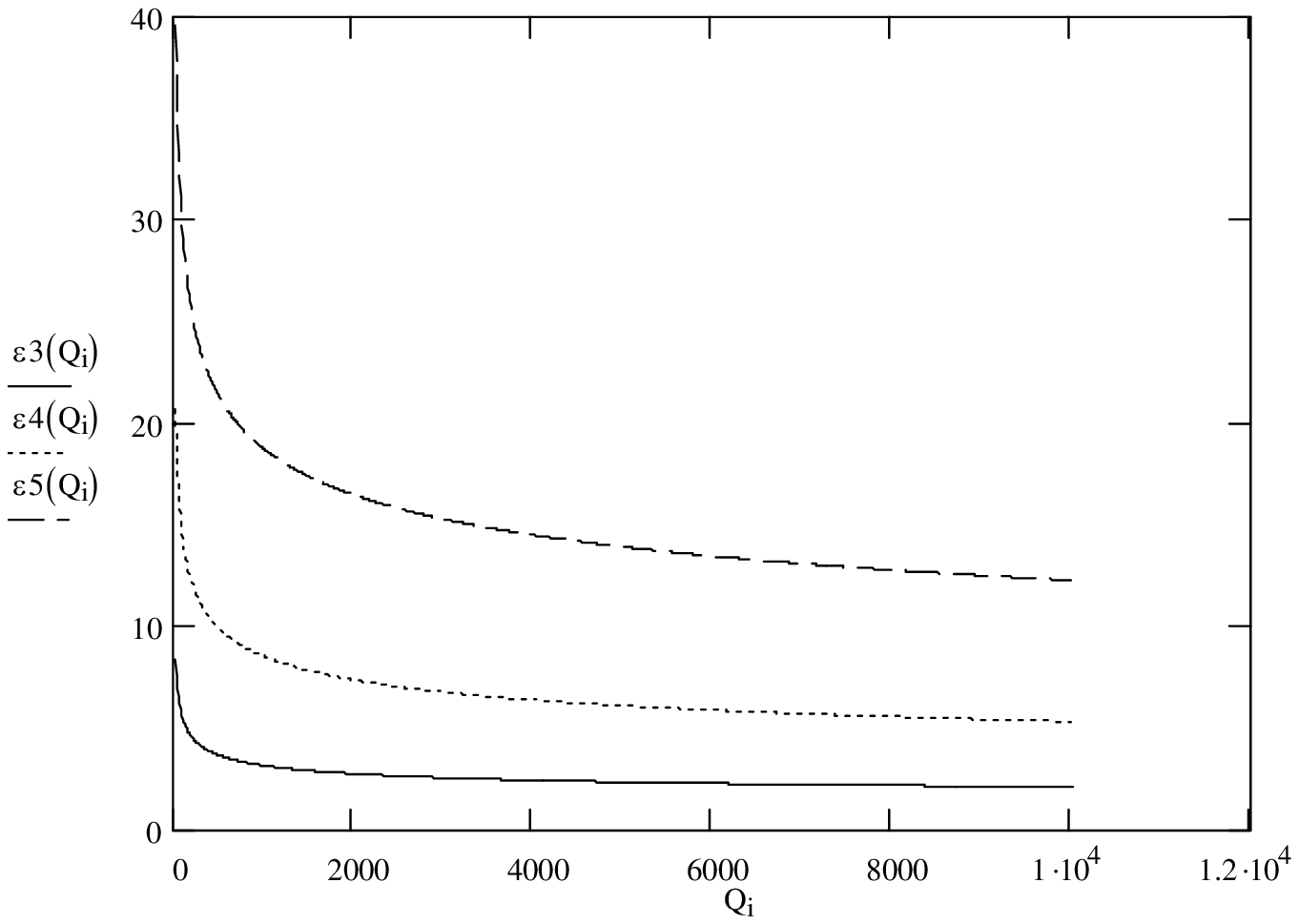}
  \caption{The solid, dot and dashed curves of the minimum energy per unit charge of large Q-balls in the complex
signum-Gordon model as functions of charge $Q$ for spatial
dimensions $d=3, 4, 5$ respectively with $\lambda=1$ and
$\Phi_{c}=1$.}
\end{figure}

\newpage
\begin{figure}
\setlength{\belowcaptionskip}{10pt} \centering
  \includegraphics[width=15cm]{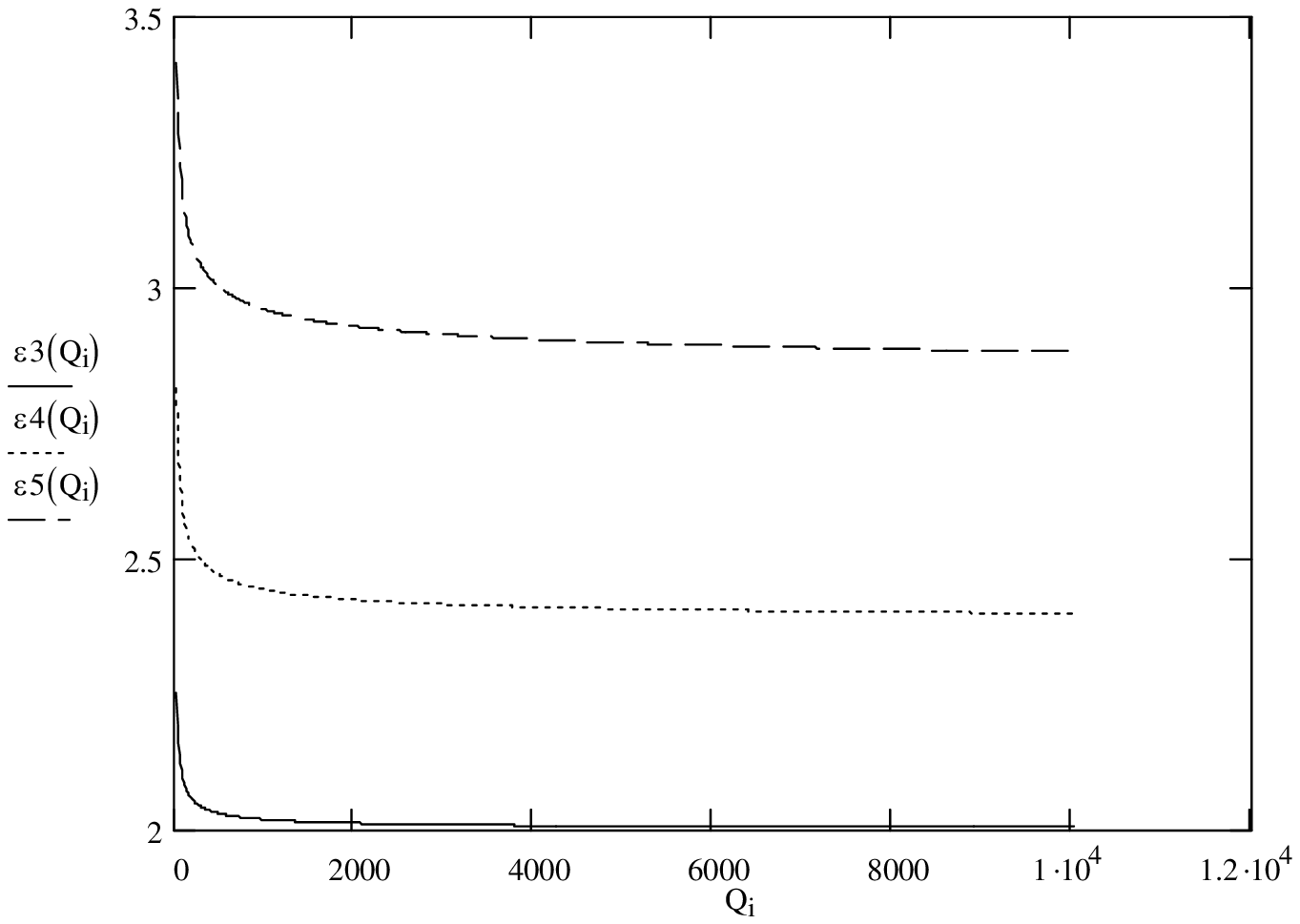}
  \caption{The solid, dot and dashed curves of the minimum energy per unit charge of small Q-balls in the complex
signum-Gordon model as functions of charge $Q$ for spatial
dimensions $d=3, 4, 5$ respectively with $\lambda=1$ and
$\Phi_{c}=1$.}
\end{figure}

\end{document}